\documentclass[11pt,a4paper]{article}

\usepackage{amsmath}
\usepackage{amsthm}
\usepackage{amssymb}
\usepackage{latexsym}
\usepackage{graphicx,subfig}
\usepackage{mathrsfs}
\usepackage{color}
\usepackage{pinlabel-test}

\allowdisplaybreaks

\newcommand{\A}{\mathcal A}
\renewcommand{\d}{\delta}

\newcommand{\e}{\varepsilon}
\newcommand{\g}{\gamma}
\newcommand{\F}{\mathscr F}

\renewcommand{\H}{\mathscr H}
\newcommand{\hf}{\mathcal H}
\renewcommand{\k}{k}
\newcommand{\kh}{\kappa_H}
\newcommand{\kg}{\kappa_G}

\renewcommand{\L}{\mathcal L}
\newcommand{\N}{\mathbb{N}}
\newcommand{\R}{\mathbb{R}}
\newcommand{\s}{\sigma}
\renewcommand{\S}{\Sigma}

\newcommand{\vfi}{\varphi}
\newcommand{\V}{\textup{Vol\,}}
\newcommand{\vs}{\textup{Vol\,}(\S)}

\newcommand{\ds}{\displaystyle}

\newcommand{\loc}{\textup{loc}}

\newcommand{\weakto}{\rightharpoonup}
\newcommand{\weaksto}{\stackrel{*}{\rightharpoonup}}

\def\XXint#1#2#3{{\setbox0=\hbox{$#1{#2#3}{\int}$}
     \vcenter{\hbox{$#2#3$}}\kern-.5\wd0}}

\newtheorem{theo}{Theorem}
\newtheorem{lemma}[theo]{Lemma}

\newtheorem{prop}[theo]{Proposition}

\newtheorem{remark}{\mdseries{\itshape{Remark}}}

\newtheorem{definition}{\bfseries{\upshape{Definition}}}
\newenvironment{defi}{\begin{definition} \upshape}{\end{definition}}

\begin {document}

\title{Global minimizers for axisymmetric multiphase membranes.}

\author{Rustum Choksi\footnote{Department of Mathematics and Statistics,
McGill University,  Montreal, Canada, {rchoksi@math.mcgill.ca}} \, \,\, \, Marco Morandotti\footnote{Department of Mathematical Sciences, Carnegie Mellon University, Pittsburgh, PA, USA, \mbox{marcomor@andrew.cmu.edu}}\, \,\, \,  Marco Veneroni\footnote{Department of Mathematics ``F. Casorati", University of Pavia, Italy, {marco.veneroni@unipv.it}}}

\date{\today}
\maketitle
\begin{abstract} 
We consider a Canham-Helfrich-type variational problem  defined over closed surfaces enclosing a fixed volume and 
having fixed surface area. The problem models the  shape of multiphase biomembranes. 
It consists of minimizing the sum of the Canham-Helfrich energy, in which  the bending rigidities and spontaneous curvatures are now phase-dependent, and a line tension penalization for the phase interfaces. By restricting attention to axisymmetric surfaces and phase distributions, we extend our previous results for a single phase \cite{ChoksiVen} and 
prove existence of a global minimizer. 
\end{abstract}
\bigskip

\noindent \textbf{Keywords:} Helfrich functional, biomembranes, global minimizers, axisymmetric surfaces, multicomponent vesicle.
\medskip

\noindent \textbf{AMS subject classification}: 49Q10, 49J45 (58E99, 53C80, 92C10).


\section{Introduction and main result}

Lipid bilayers are the most elementary and indispensable structural component of biological membranes, which form the boundary of all cells in living systems. In biological membranes, the bilayer consists of many different lipids and other amphiphiles, which can separate into coexisting liquid phases, or domains, with distinct compositions. Two types of phases are typically observed: a liquid phase with short-range order and a liquid disordered phase (see, e.g., \cite{JuelicherLipowsky96}, \cite{BaumgartHessWebb}), which we label phase $A$ and $B$ in the sequel. Their configurations, however, are fundamentally distinct from other interfaces, since they are not determined by a surface tension but rather by a bending elasticity, as introduced independently by \cite{Canham70}, \cite{Helfrich73}, and \cite{Evans74}. 


As in, e.g., \cite{JuelicherLipowsky96} and \cite{BaumgartHessWebb}, the Canham-Helfrich-Evans energy functional of a surface $\S$ and a phase $\vfi$ is given by 
\begin{equation}
\label{eq:helfrich}
	\F (\S,\vfi)=   \int_\S \left(\frac{\kh(\vfi)}{2}(H-H_0(\vfi))^2+\kg(\vfi)\, K\right) dS +\sigma \hf^1(\Gamma),
\end{equation}
where $dS$ denotes the integration with respect to the ordinary two-dimensional area measure, $H$ is the sum of the principal curvatures of $\S$, i.e., twice the mean curvature and $K$ is the Gaussian curvature. In a state where the phases are completely separated by a sharp interface, we let $\vfi:\S \to \{0,1\}$ denote the characteristic function of the phase $A$,  $\kh(\vfi),\kg(\vfi)$ are the phase-dependent bending rigidities, and $H_0(\vfi)$ is the phase-dependent spontaneous curvature. In the last term of \eqref{eq:helfrich}, $\s>0$ is the (constant) line tension coefficient, $\Gamma$ is the interface between the phases $A$ and $B$, and $\hf^1$ is the one-dimensional Hausdorff measure.

\begin{figure}[ht]
\begin{center}
\begin{minipage}{4.5cm}
\centering{
\labellist
		\hair 2pt
		\pinlabel $\S$ at 280 250
		\pinlabel $z$ at 185 370	
		\pinlabel $\g$ at 215 311
		\pinlabel $\vfi=0$ at 60 270
		\pinlabel $\vfi=1$ at 60 100
		\pinlabel $\Gamma$ at 168 202 
	\endlabellist				
\includegraphics[height=6cm]{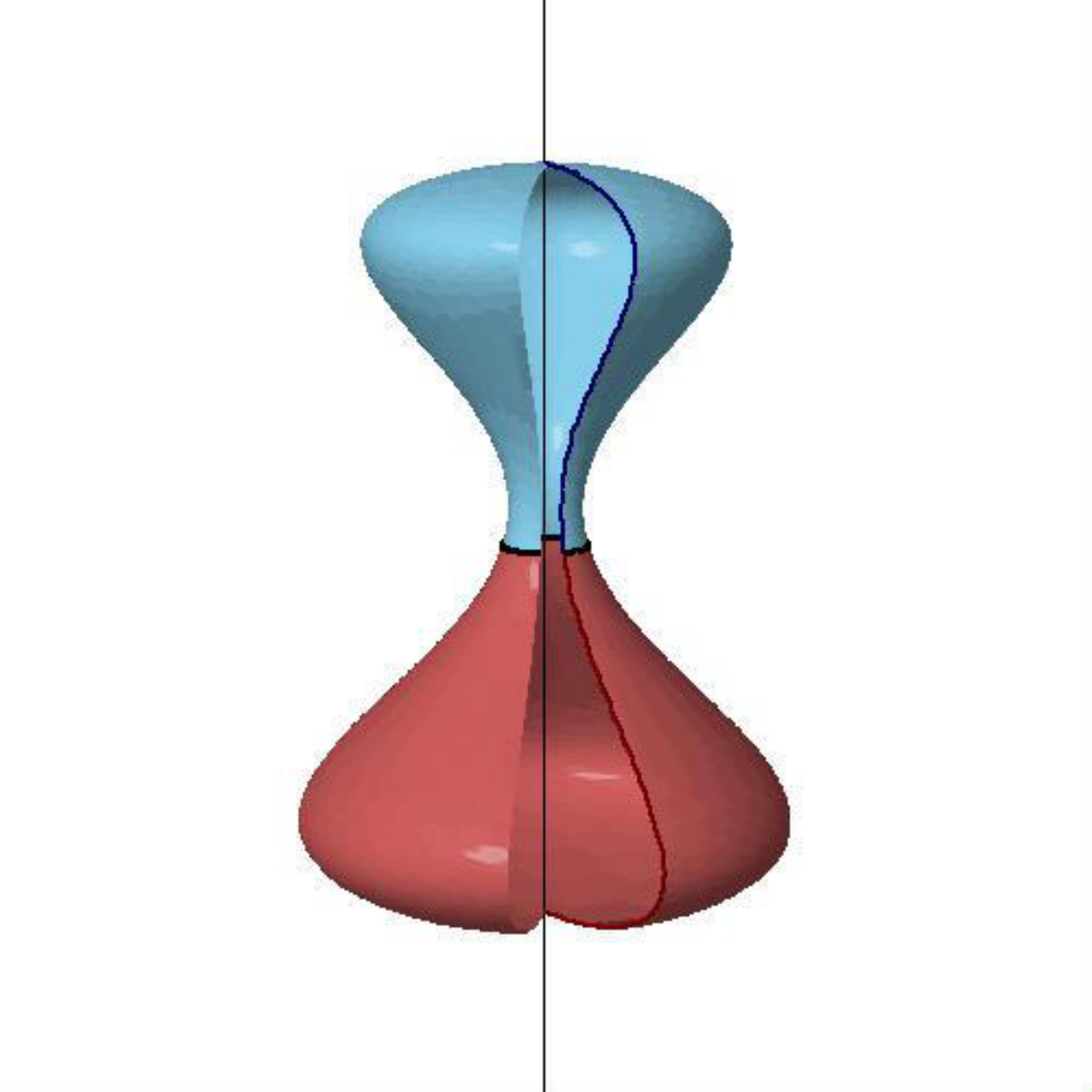}\\ 
}
\end{minipage}
\hspace{1.5cm}
\begin{minipage}{5cm}
\labellist
		\hair 2pt
		\pinlabel $\S$ at 280 250
		\pinlabel $z$ at 185 350	
	\endlabellist				
\centering{
\includegraphics[height=6cm]{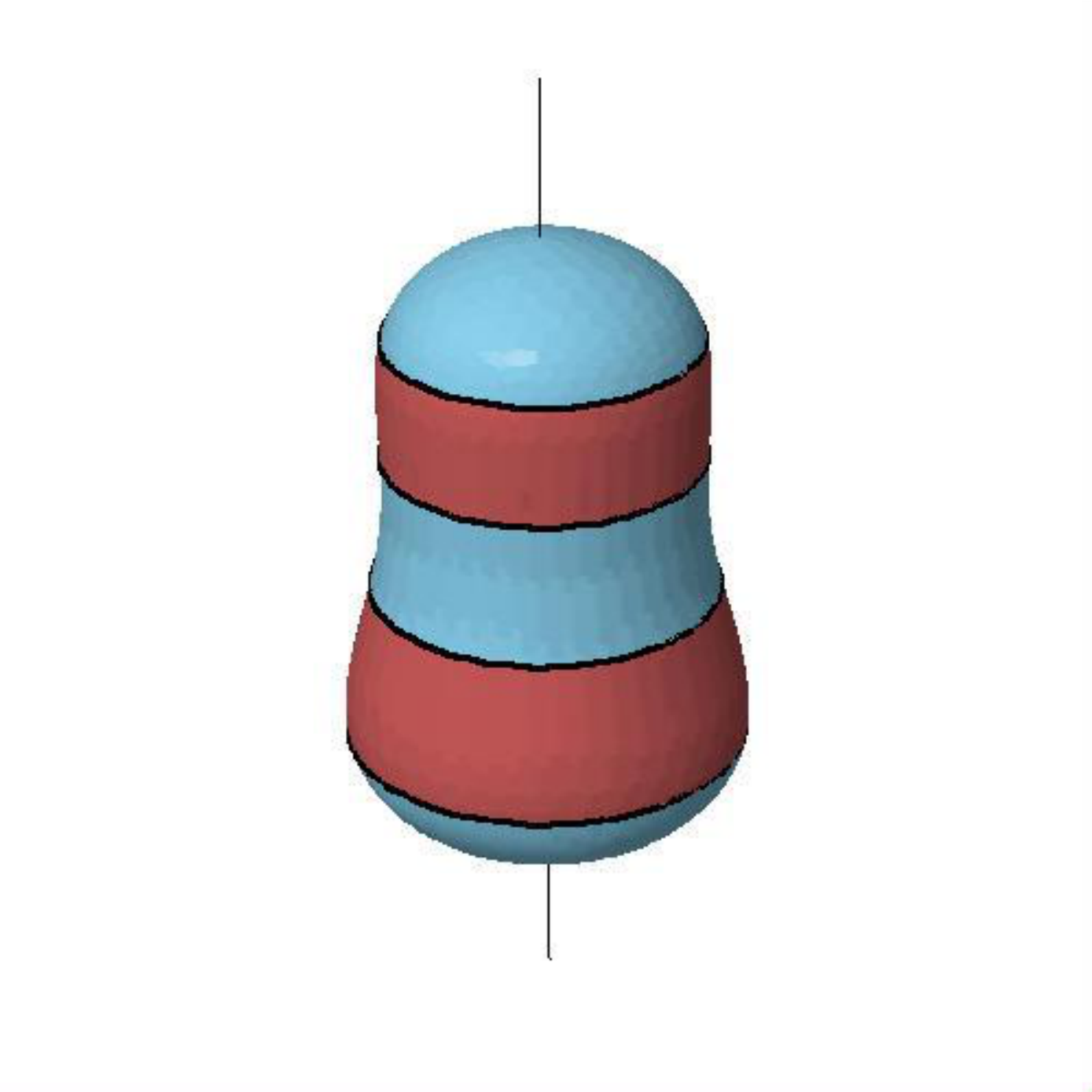}\\  
}
\end{minipage}
\end{center}
\caption{Two examples of axisymmetric multiphase surfaces.}
\label{fig2}
\end{figure}
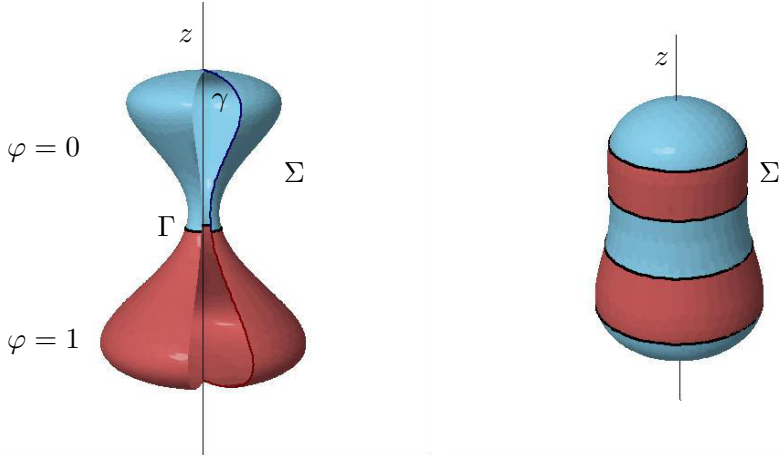


We prove the existence of a minimizer of \eqref{eq:helfrich} in the set of families of axisymmetric surface-phase couples $(\Sigma,\vfi)$  
such that the total area of the surface $\S$, the area ratio between the phases, and the volume enclosed by the surface $\S$ are fixed.
\bigskip

In the last forty years, homogeneous membranes have been extensively studied from the experimental, theoretical, and numerical points of view (see, e.g., \cite{Seifert97}). Investigation of inhomogeneous systems started more recently (\cite{JuelicherLipowsky93}, \cite{JuelicherLipowsky96}) and is nowadays at the center of an increasing focus. Multicomponent vesicles have recently been considered in numerical studies aiming at understanding the dynamics of the phase separation, the stability of nanodomains, and the complex morphology of the membranes; see the review paper \cite{Fried}. We refer in particular to the paper \cite{Sohn_etal10} concerning the dynamics of one-dimensional curves in viscous fluids, while two-dimensional surfaces where investigated with the phase field method in \cite{Du08} and \cite{ElliottStinner10P}, by means of surface finite elements in \cite{ElliottStinner10}, and by adaptive finite elements in \cite{LowengrubRaetzVoigt09}. Our result provides a theoretical basis, at least in the axisymmetric case, for the existence of the shapes approximated by these numerical studies. More advanced models, with respect to \eqref{eq:helfrich}, which couple the chemical and mechanical properties of lipid bilayers, are subject of current research, see, e.g., \cite{Zurlo}, \cite{DeseriPiccioniZurlo08}. 

We note that in the case where the spontaneous curvature vanishes, the Canham-Helfrich functional becomes the famous Willmore functional (see, e.g., \cite[Chapter 7]{Willmore93}).  While there has recently been tremendous amount of research associated with minimizing the Willmore functional, 
this research does not directly carry over to the doubly-constrained  Canham-Helfrich functional (cf. \cite{ChoksiVen}).

\subsection{Axisymmetric multicomponent vesicles}
We detail now the assumptions on the variables, the parameters, and the constraints that compose our problem and we state the main result.

\paragraph{i) Surfaces.} Axisymmetric surfaces, i.e., surfaces of revolution, can be obtained by rotating a curve about a line. 
Let $\R^2_+:=\{(x,y,z)\in \R^3 : x \geq 0, y=0\}$ be a half-plane in $\R^3$; we first consider a smooth curve $\g:[0,1]\to \R^2_+$\,, $t\mapsto (\g_1(t),0,\g_2(t))$.  By rotating $\g$ around the axis $x=y=0$, we obtain the surface $\S$ parametrized by:
\begin{equation}
\label{def:parametr}
	r(t,\theta)=\big[\g_1(t)\cos (\theta),\ \g_1(t)\sin (\theta),\ \g_2(t)\big],\qquad (t,\theta)\in [0,1]\times [0,2\pi].
\end{equation}	
If a surface $\S$ admits the parametrization \eqref{def:parametr}, then we say that $\S$ \textit{is generated by} $\g$. A standard computation (see Section \ref{sec:surrev} below) shows that if $\g$ generates a smooth surface $\S$ without boundary, then the 2-dimensional surface area $|\S|$, the enclosed volume $\vs$, and the principal curvatures $k_1,k_2$ of the generated surface are given by
\begin{align}
		 |\S| &=2\pi \int_0^1 \g_1|\dot \g|\, dt, & \vs &=\pi \int_0^1 \g_1^2\dot \g_2\, dt, \label{eq:firsdef} \\
		 k_1 &= \frac{(\ddot \g_2 \dot \g_1 -\ddot \g_1 \dot \g_2)}{|\dot \g|^3 },& k_2 &= \frac{\dot \g_2 }{\g_1|\dot \g| }. \label{eq:secdef}
\end{align}	
Denoting by $\L^1$ the one-dimensional Lebesgue measure, let $\mu_\g:= 2\pi \g_1|\dot \g|\L^1 \llcorner_{[0,1]}$ be the area measure induced by the curve $\g$. It is not difficult to see, at least in the case of only one phase (see \cite{ChoksiVen}), that a bound on $\F(\S)$ provides an a priori estimate on $\ddot \g $ in the space $L^2((0,1);\mu_\g)$, which translates into a bound for $\g$ in the space $W^{2,2}$ on any stretch of curve such that $\g_1\geq \e>0$.  Precisely, by Proposition \ref{prop:compcurv}, it is not restrictive to assume that 
\begin{equation}
\label{eq:defg}
	\g \in C^1((0,1);\R^2_+),\qquad \g_1>0\quad \forall\,t\in (0,1),\qquad \ddot\g \in L^2((0,1);\mu_\g).
\end{equation}
Moreover, in order for $\F(\S)$ to be finite,  $\g$ is forced to meet the $z-$axis orthogonally, so if $\g_1(0)=\g_1(1)=0$, then $r$ is the $C^1_\loc$ immersion of a closed surface.  If $r$ is not an embedding, as, for example, in the case of the surface generated by the limit curve in Figure \ref{fig1}, we \emph{define} the \emph{generalized} two-dimensional surface area, enclosed volume and principal curvatures of the generated surface by the quantities in \eqref{eq:firsdef}-\eqref{eq:secdef}.

\paragraph{ii) Phases.} If $\tilde \vfi$ is the characteristic function of phase $A$, then in order for $\F(\S,\tilde \vfi)$ to be bounded, and in particular in order for $\s \hf^1(\Gamma)$ to be bounded, we need to impose some kind of regularity on the class of admissible phases $\tilde \vfi:\S \to \{0,1\}$, for example by restricting to characteristic functions of finite perimeter sets on the surface $\S$. Under the simplifying assumption of  axisymmetry for the phases, as well as for the surfaces, a useful approach is then to follow the parametrization $\vfi=\tilde \vfi \circ \g:[0,1]\to \{0,1\}$. 
Let $J(\vfi)\subset(0,1)$ be the set of points where $\vfi$ has a jump discontinuity. Owing to axisymmetry, the (measure-theoretical) interface $\Gamma$ between the two phases on the surface $\S$ is a union of circles
$$\Gamma = \left\{r(t,\theta):t\in J(\vfi), \theta\in [0,2\pi] \right\}$$
and
\begin{equation}
\label{eq:defline}
	\hf^1(\Gamma) = 2\pi \int_0^1 \g_1(t)d\|D\vfi\|= 2\pi\sum_{t\in J(\vfi)}\g_1(t),
\end{equation}
where $\|D\vfi\|=\hf^0 \llcorner_{J(\vfi)}$ is the counting measure restricted to the jump set of $\vfi$ (see Section \ref{ssec:bv} below). Since $\g_1 \in C^0([0,1])$ and $\g_1(t)>0$ for $t\in (0,1)$, then according to \eqref{eq:defline}, $\hf^1(\Gamma)<+\infty$ if and only if 
$\g_1\in L^1(\|D\vfi\|)$. In the sequel, it will be convenient to deal with the weaker request that 
\begin{equation}
\label{eq:phase}
	\vfi \in BV_\loc((0,1);\{0,1\}),
\end{equation}
extending $\F$ to $+\infty$ if the quantity in \eqref{eq:defline} is not bounded. The area measure of phase $A$ can then be expressed as
\begin{equation}
\label{eq:areaA}
	\int_\S \tilde \vfi\, dS =  2\pi \int_0^1 \vfi (t) \g_1(t)|\dot\g|\, dt.
\end{equation}
It would also be possible to choose, as ambient space for the phases, the space of \emph{special} functions of (locally) bounded variation $SBV_\loc$\,. Since we are dealing with two-valued functions, and the Cantorian part of the measure will not appear in any case, we prefer to keep the setting as simple as possible and use $BV_\loc$ functions.  

\paragraph{iii) Parameters.} Let the bending rigidities for phase $i$ be given by $\kh^i\,,\kg^i$\,, and the spontaneous curvature be $H_0^i$\,, for $i=A,B$. In order for the the functional $\F$ to be coercive, we require (see Lemma \ref{lemma:fundest} below) that
\begin{equation}
\label{eq:assk}
	\kh^i>0 \quad \text{and}\quad  \frac{\kg^i}{\kh^i} \in (-2,0) \quad \text{for }i=A,B.
\end{equation}
We note that  the physical range in which the parameters $\kh$ and $\kg$ are typically found is contained in the one we impose in \eqref{eq:assk}, see e.g. \cite{TemplerKhooSeddon} and \cite{BaumgartDWJ} (note that the latter cites the former, but inverting numerator and denominator, by mistake).

We define the coefficient functions on the interval $[0,1]$ in such a way that $\kh$ is the linear interpolation of $\kh(0)=\kh^B$\,, $\kh(1)=\kh^A$\,, and the same holds for $\kg$ and $H_0$\,, i.e.,
$$ \kh(\vfi):=\vfi\kh^A +(1-\vfi)\kh^B,\quad \kg(\vfi):=\vfi\kg^A +(1-\vfi)\kg^B,\quad H_0(\vfi):=\vfi H_0^A+(1-\vfi)H_0^B.$$
According to the parametrization \eqref{def:parametr}, the contributions in \eqref{eq:helfrich} depending on the mean curvature, the Gaussian curvature, and the line tension of the interface between the phases, can then be written as
\begin{align}
		\frac 12\int_\S \kh(\vfi)(H-H_0(\vfi))^2 dS &=  \pi  \int_0^1 \kh(\vfi)\left( k_1+k_2  -H_0(\vfi)\right)^2 \g_1|\dot \g |\,dt,\label{eq:f1}\\
		\int_\S \kg(\vfi)K dS &=  2\pi  \int_0^1 \kg(\vfi) k_1 k_2 \g_1|\dot \g |\,dt,\label{eq:f2}\\
		\sigma \hf^1(\Gamma) &= 2\s \pi\int_0^1 \g_1(t)d\|D\vfi\|.\label{eq:f3} 
\end{align}


\begin{defi}
We say that a couple of  surface and phase $(\S,\vfi)$ is \emph{admissible} if the surface $\S$ is generated by a curve $\g$ satisfying \eqref{eq:defg} and the phase $\vfi$ satisfies \eqref{eq:phase}. 
\end{defi}

Finally, for fixed area and volume, a configuration made of several connected components may have a lower energy than a one-component configuration. This could also be favored by a relatively high value of $\s$, since separation of phases in different components would have no interface between phases and thus $\s\hf^1(\Gamma)=0$. From a dynamical point of view, in certain conditions, shape transformations involving topological changes, like budding and fission (see, e.g., \cite[Section 3]{Seifert97}), could be expected. We take this possibility into account by studying families of admissible surface-phase couples $S=\left\{(\S_1,\vfi_1),\ldots,(\S_m,\vfi_m)\right\}$, and defining the total energy of such a system as the sum of the Helfrich energies of the single components: $\mathcal F(S):= \sum_{i=1}^m \F(\S_i,\vfi_i)$.

\paragraph{iv) Main result.} We prove the following.

\begin{theo}
\label{th:main}
Let $\kh^i,\kg^i\in \R$, $i=A,B$ be given such that \eqref{eq:assk} is satisfied. Let $\s>0,$ $H_0^i\in \R$, $i=A,B$ be given. Let $\A,\Pi_A,\mathcal V>0$ be given such that 
\begin{equation}
\label{eq:apiv}
	\mathcal V< \frac{{\A}^{3/2}}{6 \sqrt{\pi}},\qquad 0\leq \Pi_A\leq \A.
\end{equation}
Let $\mathcal S(\A,\Pi_A,\mathcal V)$ denote the set of finite families $S=\left\{(\S_1,\vfi_1),\ldots,(\S_m,\vfi_m)\right\}$ of admissible couples of surfaces and phases such that the \emph{generalized} area, volume, and phase area constraints
$$\sum_{i=1}^m|\S_i|=A,\qquad \sum_{i=1}^m \V(\S_i)=V,\qquad \sum_{i=1}^m 2\pi\int_0^1\vfi_i(t)\g_1(t)|\dot \g|dt=\Pi_A$$
are satisfied (see \eqref{eq:firsdef},\eqref{eq:areaA}). 
Let  $\ds \mathcal F(S):= \sum_{i=1}^m \F(\S_i,\vfi_i)$. Then the problem
$$ \min \left\{\mathcal F(S): S\in \mathcal S(\A,\Pi_A,\mathcal V)\right\}$$
has a solution.
\end{theo}

\subsection{Structure of the proof and plan of the paper }
The proof of Theorem \ref{th:main} follows the direct method of the calculus of variations: given a minimizing sequence of (systems of) surfaces and phases $S^n$ satisfying the area and volume constraints, by compactness we obtain a subsequence converging to a system $S$, and by lower semicontinuity of the functional $\mathcal F$ we prove that $S$ is a global minimizer. 

Regarding compactness, it is fundamental that the phase-dependent parameters $\kh$\,, $\kg$\,, $H_0$ are chosen in such a way that the functional $\F$ is an upper bound for the $L^2$-norm of the second fundamental form of the surface $\S$ (Lemma \ref{lemma:fundest}). In this way we can exploit the compactness result obtained in \cite{ChoksiVen} in the case of homogeneous membranes. The main idea in modeling the phases (see also Section \ref{ssec:discussion}) is to follow the parametrizations $\g$ instead of their images, so that the sequence of parametrized phases is defined on a fixed interval rather than a sequence of surfaces. The drawback  with this approach is that the phases are only of \emph{locally} bounded variation; where the curves touch the axis of revolution, the area measure vanishes. In other words, where the horizontal component $\g_1$ of the curve $\g$ becomes infinitesimal, the line tension part of the functional \eqref{eq:f3} may allow for an infinite number of discontinuities in the phases $\vfi$. However, this will not constitute a problem in the proof of lower semicontinuity, since the combined phases and curves are well-behaved (Lemma \ref{lemma:hlsc}).

The proof of the lower semicontinuity for the curvature terms \eqref{eq:f1} and \eqref{eq:f2} requires a special care, since we have to pass to the limit simultaneously in the surfaces and in the phases defined on the surfaces. A useful tool can be found in the function-measure pairs introduced in \cite{Hutchinson86}.
\medskip

In Section \ref{sec:geoineq} we describe the notation,  we derive the geometrical quantities involved in Helfrich's functional,  and we recall the basic definitions and the main results regarding functions of bounded variation and measure-function couples. In Section \ref{sec:existence} we study the compactness and lower semicontinuity of a bounded admissible sequence, and end with the proof of Theorem \ref{th:main}.

\subsection{Discussion}
\label{ssec:discussion}
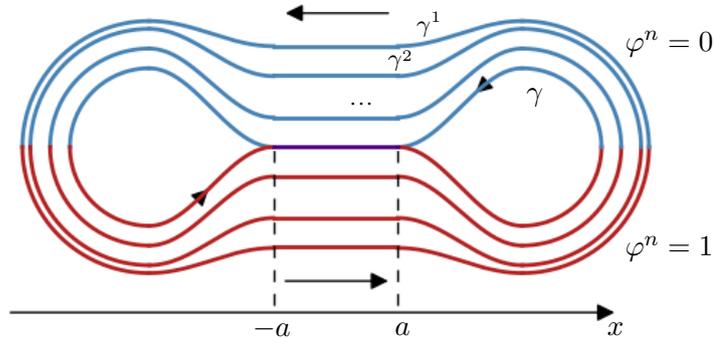
\begin{figure}[h!]
\begin{center}
	\labellist
	\hair 2pt
	\pinlabel {\footnotesize $\g^1$} at 251 176
	\pinlabel {\footnotesize $\g$} at 232 154
	\pinlabel {\footnotesize $\phantom{\g}^2$} at 235 154
	\pinlabel {\footnotesize $\cdot\! \cdot\! \cdot$} at 213 131			
	\pinlabel $\g$ at 310 135
	\pinlabel $-a$ at 163 5	
	\pinlabel $a$ at 236 5	
	\pinlabel $x$ at 355 5		
	\pinlabel $\vfi^n=0$ at 385 165	
	\pinlabel $\vfi^n=1$ at 385 50				
	\endlabellist			
\centering
\noindent
\includegraphics[height=5.4cm]{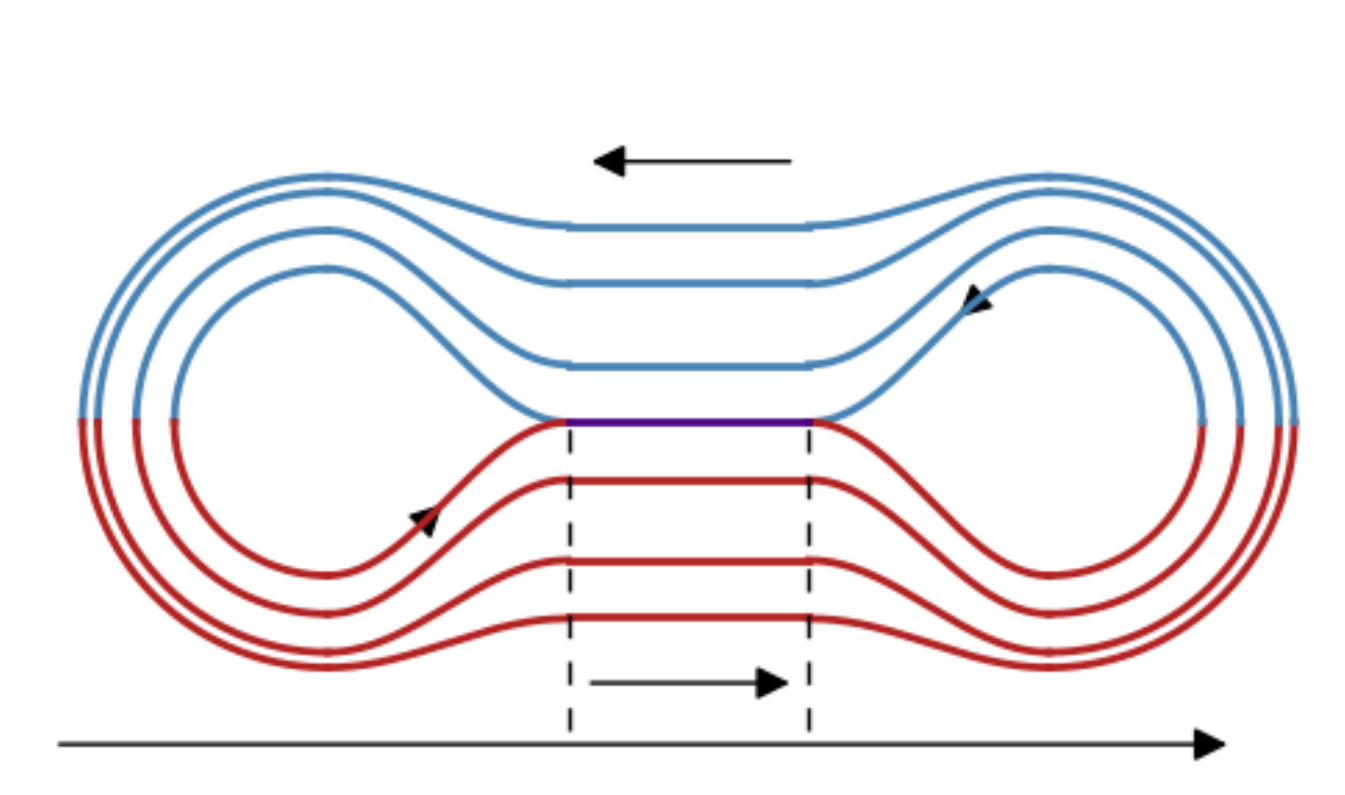}
\end{center}
\vspace{-0.5cm}

\caption{The problem of defining the phase on a curve when overlapping can arise as limit of well-defined configurations.}
\label{fig1}
\end{figure}

\paragraph{Modeling the phases.} The choice of modeling the phases by following the parametrization $\vfi=\tilde \vfi \circ \g:[0,1]\to \{0,1\}$, instead that by functions with support on the surface $\tilde \vfi:\S\to\{0,1\}$, yields first of all a much simpler setting, since the domain becomes a fixed real interval, instead of surface. It is also a way to solve a more intrinsic question related to the possible ill-posedness of the problem. Consider, as in the example in Figure \ref{fig1}, a sequence of curves $\g^n$ and phases $\vfi^n$ such that, in the limit, two stretches of curve carrying different phases overlap. How should the limit phase be defined in this case?

Any $\{0,1\}$-valued phase $\tilde \vfi$ defined on the support of $\g$ would not be a limit (not even in the sense of distributions) of the sequence $\tilde \vfi^n$. Instead, defining the parametrized phases on the interval $[0,1]$, the situation in Figure \ref{fig1} could be described, e.g., by choosing a constant sequence $\vfi^n$ equal to the characteristic function of the interval $(0,1/2)$, which clearly converges to the same characteristic function.

Another way to approach the problem of overlapping curves could be, for example, to consider the varifolds \cite{Hutchinson86} associated with the curves and to describe the segment $[-a,a]\times \{0\}$ by a stretch with double density. In the case of a homogeneous curve in two dimensions, this approach was followed, for example, in \cite{BellettiniMugnai07}, but in the case of multiple phases it is not evident which approach to follow.
A possibility could be by multivalued functions on varifolds, but we deem more natural to follow the parametrization of the curves. In higher dimensions this could be generalized by following the immersion of a Riemann surface as opposed to studying its image.
  
\paragraph{Generalizations - surfaces.} Experimental evidence of higher-genus membranes is known for homogeneous membranes (see, e.g., \cite[pages 19, 20]{Seifert97}), but we are not aware of any experiment or simulation showing multiphase biological or artificial membrane of genus higher than zero. Nonetheless, from a mathematical viewpoint, it is straightforward to allow for genus-1 axisymmetric membranes (see \cite{ChoksiVen}) by considering closed generating curves with strictly positive first component. Since the $W^{2,2}$-norm of such curves is bounded from above by \eqref{eq:helfrich}, they have the same compactness properties of the curves that we consider, and Theorem \ref{th:main} could be directly extended to include also genus-1 generators. What is not clear to us, is whether the genus of the minimizer could be prescribed, since a priori a sequence of genus-1 minimizer may degenerate to a genus-0 surface in the limit. 

\paragraph{Generalizations - phases.} Even though, according to experiments and simulations, we expect the phases of an axisymmetric surface to be axisymmetric as well, one could pose the problem of studying general phases on surfaces of revolution. We believe that this step could be performed following the same steps as in the symmetric case, as the theorems used for compactness and semicontinuity for functions of bounded variation do not depend on the dimension. Actually, the functional \eqref{eq:helfrich} for a two-dimensional phase would provide a better bound, namely in $BV$ instead of $BV_\loc$\,.

Generalizations to non axisymmetric surfaces and phases would require a completely different approach: our method relies on Ascoli-Arzel\`a compactness for equicontinuous curves, which in one dimension can be applied owing to the compact immersion $W^{1,\infty} \hookrightarrow W^{2,2}$, which fails in higher dimensions.

\paragraph{Necessity of constraints.} In order to model realistic configurations of multiphase membranes, we considered classes of surfaces with fixed total area, phases area, and enclosed volume. We notice, though, that the only constraint which is necessary in order to obtain compactness of a minimizing sequence is that on the total area, which is needed in order to bound from above the full second fundamental form of the surface (Lemma \ref{lemma:fundest}) and to ensure that the limit of a minimizing sequence does not vanish. The area of each phase is then bounded by the total area, and the enclosed volume results bounded by isoperimetric inequality. Theorem \ref{th:main} could then be restated by saying that if the total area is bounded from above and away from zero, then there is a minimizer of \eqref{eq:helfrich}, and, in particular, there is one for any admissible choice of volume and phase area constraints. 

\section{Preliminaries}
\label{sec:geoineq}

\subsection{Notation}
\label{ssec:notation}
We refer to \cite{AFP} for the definitions of the following objects. Let $\mathcal L^d$ and $\hf^d$ be the $d$-dimensional Lebesgue and Hausdorff measures, let $\mathcal L^d \llcorner_A$ and $\hf^d \llcorner_A$ be the restrictions to the set $A$ of the $d$-dimensional Lebesgue and Hausdorff measures, respectively. For an open set $E\subset \R$, let $C_c(E)$ be the space of continuous functions with compact support in $E$, let $C_0(E)$ be the closure of $C_c(E)$ with respect to uniform convergence, and let $RM(E)$ be the space of Radon measures on $E$, which can be identified with the dual of $C_c(E)$. If $\mu \in RM(E)$ and $\mu(E)<\infty,$ we say that $\mu$ is a finite Radon measure.  
For a curve $\g:[0,1]\to \R^2$, $t\mapsto (\g_1(t),\g_2(t))$, we use the shorthand notation $\{\g_1 \geq 0\}$ to denote the set  $\{t\in [0,1]:\g_1(t) \geq 0\}$, we denote the first and second derivatives by $\dot \g, \ddot \g$, and we define the measure $\mu_\g\in RM(\R)$ by $\mu_\g:= 2\pi \g_1|\dot \g|\L^1 \llcorner_{[0,1]}$.

\subsection{Parametrization}
\label{sec:surrev}

The derivation of the principal curvatures for a surface of revolution can be found, e.g., in \cite[Section 3-3, Example 4]{doCarmo} or in \cite{ChoksiVen}. Recalling the parametrization introduced in \eqref{def:parametr}, we define the tangents to the surface along the coordinate lines as
\begin{align*}
	r_t:=\frac{\partial}{\partial t}r(t,\theta)&= \big[\dot \g_1(t)\cos \theta,\ \dot\g_1(t)\sin \theta,\ \dot\g_2(t)\big], \\
	r_\theta:=\frac{\partial}{\partial \theta}r(t,\theta)&= \big[-\g_1(t)\sin \theta,\ \g_1(t)\cos \theta,\ 0\big],
\end{align*}
and notice that $r_t\cdot r_\theta=0$, i.e., they are always orthogonal to each other. The first fundamental form is given by
$$	g(t,\theta)=\left[
	\begin{array}{cc}
		E & F\\
		F & G
	\end{array}	
	\right]=
	\left[
	\begin{array}{cc}
		r_t \cdot r_t & r_t \cdot r_\theta\\
		r_\theta \cdot r_t & r_\theta \cdot r_\theta
	\end{array}	
	\right]
	=\left[
	\begin{array}{cc}
		|\dot \g(t)|^2 & 0 \\
 		0		& \g_1(t)^2
	\end{array}	
	\right],
$$	
$$ \sqrt{ g} := \sqrt{\det (g_{ij})}=\g_1(t)|\dot \g(t)|.$$
The normal vector to the surface can be oriented either inwards or outwards, depending on the direction of $\g$. For curves parametrized in counterclockwise direction, it is inwards:
\begin{align*}
	n(t,\theta) = \frac{r_t \times r_\theta}{\sqrt g}&=\frac{1}{\g_1(t)|\dot \g(t)|}\big[ -\g_1(t)\dot \g_2(t) \cos \theta,-\g_1(t)\dot \g_2(t) \sin \theta,\ \g_1(t)\dot \g_1(t)\big]\\
		&= \frac{1}{|\dot \g(t)|}\big[ -\dot \g_2(t) \cos \theta,-\dot \g_2(t) \sin \theta,\ \dot \g_1(t)\big].
\end{align*}		
Now we start to compute the second fundamental form, using a constant-speed parametrization.
\begin{align*}
	n_t:= \frac{\partial}{\partial t}n(t,\theta) &=\frac{1}{|\dot \g(t)|}  \big[ -\ddot \g_2(t) \cos \theta,-\ddot \g_2(t) \sin \theta,\ \ddot \g_1(t)\big]\\
	n_\theta:= \frac{\partial}{\partial \theta}n(t,\theta) &=\frac{1}{|\dot \g(t)|} \big[ \dot \g_2(t) \sin \theta,-\dot \g_2(t) \cos \theta,\ 0\big].
\end{align*}
The second fundamental form is given by
$$	II(t,\theta)=\left[
	\begin{array}{cc}
		L & M\\
		M & N
	\end{array}	
	\right]=-\left[
	\begin{array}{cc}
		n_t \cdot r_t & n_t \cdot r_\theta\\
		n_\theta \cdot r_t & n_\theta \cdot r_\theta
	\end{array}	
	\right]
	=\frac{1}{|\dot \g(t)|} \left[
	\begin{array}{cc}
		\ddot \g_2 \dot \g_1 -\ddot \g_1 \dot \g_2  & 0 \\
 		0		& \g_1 \dot \g_2
	\end{array}	
	\right].
$$	
The \textit{signed} curvature of $\g$ is defined as
$$
 	\k:=\frac{\ddot \g_2 \dot \g_1 -\ddot \g_1 \dot \g_2}{|\dot \g|^3}.
$$
The Gaussian curvature is given by
$$ K=k_1 k_2 = \frac{LN -M^2}{EG - F^2}=\frac{(\g_1 \dot \g_2)(\ddot \g_2 \dot \g_1 -\ddot \g_1 \dot \g_2)}{(\g_1)^2|\dot \g|^4 }=\frac{\dot \g_2 \k}{\g_1|\dot \g|}.$$
The mean curvature is (for sake of notation, we define as mean curvature the double of what is often defined as mean curvature)
\begin{equation*}
	H=k_1 +k_2 = \frac{LG-2MF +NE }{EG - F^2}=\frac{\g_1^2 (\ddot \g_2 \dot \g_1 -\ddot \g_1 \dot \g_2) + \g_1\dot \g_2 |\dot \g|^2}{(\g_1)^2|\dot \g|^3 }
		= \k + \frac{\dot \g_2}{\g_1|\dot \g|} .\\
\end{equation*}
The principal curvatures are
\begin{align*}
	k_1&=\k=\frac{\ddot \g_2 \dot \g_1 -\ddot \g_1 \dot \g_2}{|\dot \g|^3}, 		\qquad \text{(meridian)}&&
	k_2=\frac{\dot \g_2}{\g_1|\dot \g| }.\qquad \text{(parallel)}
\end{align*}	
The area is
\begin{equation}
\label{eq:defarea}
	|\S|=\int_\S dS=\int_0^{2\pi}\int_0^1 \sqrt{g(t)}\,dt\,ds= 2\pi \int_0^1 \g_1(t)\, |\dot \g(t)|\, dt= \int_0^1 d\mu_\g,
\end{equation}
and the volume enclosed by the surface is
$$
	\vs=\pi \int_0^1 \g_1^2(t)\dot \g_2(t)\,  dt.
$$
It is then straightforward to check that Helfrich energy for axisymmetric surface and phase is given by expressions \eqref{eq:f1}--\eqref{eq:f3}.

\subsection{Functions of bounded variation}
\label{ssec:bv}
Let $U$ denote an open subset of $\R$; following \cite{EvansGariepy92} we say that a function $f\in L^1(U)$ has bounded variation in $U$, and write $f\in BV(U)$, if
$$ \sup\left\{\int_U f\psi'\, dx : \psi \in C^1_c(U),|\psi|\leq 1  \right\}<\infty.$$
We say that a function $f\in L_\loc^1(U)$ has locally bounded variation in $U$, and write $f\in BV_\loc(U)$, if for each open set $V \subset \subset U$
$$ \sup\left\{\int_V f\psi'\, dx : \psi \in C^1_c(V),|\psi|\leq 1  \right\}<\infty.$$
If $f\in BV_\loc(U)$ there exists a finite Radon measure $\mu$ and a $\mu$-measurable function $\eta:U\to\R$ such that
$$ |\eta|=1\quad \mu\,\text{-a.\,e., and }\quad \int_U f \psi'\, dx = -\int_U \psi\,\eta\,d\mu$$
for all $\psi \in C^1_c(V)$. We write $\|D f\|=\mu$ and $[Df]=\eta\|Df\|$. If $f\in BV(U)$, then for each $V \subseteq U$
$$\|Df\|(V)= \sup\left\{\int_V f\psi'\, dx : \psi \in C^1_c(V),|\psi|\leq 1  \right\}<\infty.$$
(Since $Df$ is scalar, we should have written $|Df|$ instead of $\|Df\|$, but $|Df|$ and $[Df]$ look too similar). For example, denoting by $\delta_x$ the Dirac distribution centered at $x$, if $f\in BV(U)$ is the characteristic function of an interval $(a,b)\subset U$, then $[Df]=\delta_a-\delta_b$\,, $\|Df\|=\d_a+\d_b$\,,
$$ \int_U f \psi'\, dx = -\int_U \psi\,d[Df]=\psi(b)-\psi(a),\qquad \int_U \psi\,d\|Df\|=\psi(b)+\psi(a)$$
and $\|Df\|(U)=\hf^0\{a,b\}=2$.

In the sequel we will rely on the two fundamental results (\cite[Section 5.2]{EvansGariepy92})
\begin{theo}[Lower Semicontinuity in $BV$]
\label{th:thlsc}
Let $f^k,f \in BV(U)$ be such that $f^k\to f$ in $L^1_\loc(U)$. Then
$$ \liminf_{k\to \infty} \|Df^k\|(U)\geq\|Df\|(U).$$
\end{theo}
\begin{theo}[Compactness in $BV$]
\label{th:thcpt}
Let $f^k \in BV(U)$ be such that $\sup_k\{\|f^k\|_{L^1(U)}+\|Df^k\|(U)\}<\infty$. Then there exist a subsequence $f^{k_j}$ and a function $f\in BV(U)$ such that $f^{k_j}\to f$ in $L^1(U)$.
\end{theo}


\subsection{A notion of convergence for measure-function couples}
\label{ssec:mt}
We recall that a sequence of Radon measures $\mu^n$ is said to converge weakly-$*$ to $\mu\in RM(\R)$ if
$$ \lim_{n\to \infty} \int_\R  \phi(t)\, d\mu^n(t) \to  \int_\R  \phi(t)\, d\mu(t)$$
for every $\phi \in C_c(\R)$.
We define the space of $p$-summable functions with respect to a positive Radon measure $\mu$ as
$$ L^p(\mu;\R^2):=\left\{ f:\R\to \R^2\ \mu\text{-measurable, such that }\int_\R |f(x)|^p\, d\mu(x)<+\infty \right\}.$$

\begin{defi}[\textit{Convergence of measure-function couples}]
\label{def:weakags}
Following \cite[Definition 5.4.3]{AGS}, given a sequence of measures $\mu^n\in RM(\R)$ converging weakly-$*$ to $\mu$, we say that a sequence of (vector) functions $f^n\in L^1(\mu^n;\R^2)$ converges weakly to a function $f\in L^1(\mu;\R^2)$, and we write $f^n \weakto f$ in $L^1(\mu^n;\R^2)$, provided
\begin{equation}
\label{eq:weakags}
	\lim_{n\to \infty} \int_\R f^n (t)\cdot \phi(t)\, d\mu^n(t) \to  \int_\R f (t)\cdot \phi(t)\, d\mu(t)
\end{equation}
for every $\phi \in C^\infty_c(\R;\R^2)$. For $p>1$, we say that a sequence of (vector) functions $f^n\in L^p(\mu^n;\R^2)$ converges \emph{weakly} to a function $f\in L^p(\mu;\R^2)$, and we write $f^n \weakto f$ in $L^p(\mu^n;\R^2)$, provided
\begin{equation}
\label{eq:weakagsp}
	\sup_{n\in \N} \int_\R |f^n (t)|^p\, d\mu^n(t) <+\infty\quad \text{and}\quad f^n \weakto f \text{ in }L^1(\mu^n;\R^2).
\end{equation}
For $p>1$, we say that a sequence of (vector) functions $f^n\in L^p(\mu^n;\R^2)$ converges \emph{strongly} to a function $f\in L^p(\mu;\R^2)$, and we write $f^n \to f$ in $L^p(\mu^n;\R^2)$, if \eqref{eq:weakagsp} holds and
$$ \limsup_{n \to \infty} {\|f^n\|}_{L^p(\mu^n;\R^2)}\leq {\|f\|}_{L^p(\mu;\R^2)}\,.$$
\end{defi}
\begin{lemma}[Weak-strong convergence in $L^p(\mu;\R^d)$ {\cite[Proposition 3.2]{Moser01}}]
\label{lemma:moser}
Let $p,q\in (1,\infty)$ such that $1/p+1/q=1$. Suppose that $\mu^n$ and $\mu$ are Radon measures on $\R$ and that $f^n\in L^p(\mu^n;\R^d)$, $f\in L^p(\mu;\R^d)$, $g^n\in L^q(\mu^n;\R^d)$, $g\in L^q(\mu;\R^d)$ be such that
$$
	f^n \weakto f\quad \text{weakly in }L^p(\mu^n;\R^d),\qquad 	g^n \to g\quad \text{strongly in }L^q(\mu^n;\R^d).
$$
Then
$$
	f^n g^n \weakto fg \quad \text{weakly in }L^1(\mu^n;\R^d).
$$
\end{lemma}
\begin{theo}[Lower semicontinuity {\cite[Theorem 5.4.4 - (ii)]{AGS}}]
\label{th:ags}
Let $p>1$, let $f^n\in L^p(\mu^n;\R^2)$ be a sequence converging \emph{weakly} to a function $f\in L^p(\mu;\R^2)$ in the sense of Definition \ref{def:weakags}, then
$$
	\liminf_{n \to \infty} \int_\R g(f^n(t))\, d\mu^n(t) \geq \int_\R g(f(t))\, d\mu(t),
$$
for every convex and lower semicontinuous function $g:\R\to (-\infty,+\infty].$
\end{theo}

\section{Existence of a minimizer}
\label{sec:existence}

\setcounter{equation}{0}
\setcounter{theo}{0}

The proof of existence relies on the following fundamental estimate, which shows that, for properly chosen coefficients, the functional $\F(\S,\vfi)$ bounds from above the $L^2$-norm of the second fundamental form of $\S$, independently of the phase $\vfi$. 
Whereas the estimate follows from the phase-independent case (see \cite[Lemma 2.1]{ChoksiVen}),  for sake of completeness we include the details.
\begin{lemma}[Fundamental estimate]
\label{lemma:fundest}
If $\kh^i\,,\kg^i$\,, $i=A,B$ satisfy
\begin{equation}
\label{eq:boundkunif}
	\kh^i>0 \quad \text{and}\quad  \frac{\kg^i}{\kh^i} \in (-2,0) \quad \text{for }i=A,B,
\end{equation}
then there exists $C>0$ such that
$$ \F(\S,\vfi)\geq C\left( \int_\S (k_1^2 +k_2^2 )dS -|\S|  \right)$$
for all admissible couples $(\S,\vfi)$.
\end{lemma}
\begin{proof}
Let $\lambda,H_0 \in \R$, note that
\begin{equation*}
	\frac12 (k_1 +k_2)^2 +\lambda k_1k_2 = \frac12 (k_1^2 +k_2^2) + (1+\lambda)k_1k_2 \geq \frac{1-|1+\lambda|}{2}(k_1^2+k_2^2),
\end{equation*}
and the last term is positive if and only if $\lambda \in (-2,0)$. For all $\e>0$ it holds
$$
	\frac{H^2}{2} = \frac{(H-H_0+H_0)^2}{2}\leq \frac {1+\e}2 (H-H_0)^2 + \frac{1+\e}{2\e}H_0^2,
$$
and thus
$$
	\frac {1+\e}2 (H-H_0)^2 + \frac{1+\e}{2\e}H_0^2 +\lambda(1+\e) K \geq \frac{1-|1+\lambda(1+\e)|}{2}(k_1^2+k_2^2).
$$
For all $\kh>0$, choosing $\lambda = \kg/\kh \in (-2,0)$ and $\e>0$ such that $(1+\e)\kg/\kh \in (-2,0)$, we get
$$
	\frac{\kh}{2}(H-H_0)^2 + \kg K + c_1 H_0^2 \geq c_2(k_1^2+k_2^2),
$$
where $c_1=\kh/2\e$ and $c_2=\frac{\kh-|\kh+\kg(1+\e)|}{2(1+\e)}>0$. Denoting by $c_1^i\,,c_2^i$ ($i=A,B$) the constants obtained for $\kh=\kh^i$\,, $\kg=\kg^i$\,, etc., we integrate on $\S$ to obtain
\begin{equation}
\begin{split} 
	\F(\S,\vfi)&=  \int_\S \left(\frac{\kh(\vfi)}{2}(H-H_0(\vfi))^2+\kg(\vfi)\, K\right) dS +\sigma \hf^1(\Gamma) \\
			&\geq \min\left\{c_2^A,c_2^B\right\}\int_\S (k_1^2+k_2^2)dS - |\S| \max\left\{c_1^A(H_0^A)^2,c_1^B(H_0^B)^2\right\}.
\end{split}
\end{equation}
\end{proof}

\subsection{Compactness}
Let the total area, the $A$-phase area and the volume constraints $\A,\Pi_A,\mathcal V$ be given. Let 
$$S^n=\left\{(\S^n_1,\vfi^n_1),\ldots,(\S^n_{m(n)},\vfi^n_{(m(n))})\right\},$$ 
where $m(n)$ is the cardinality of the $n$-th family, be a minimizing sequence for $\F$, i.e.,  
$$ S^n\in \mathcal S(\A, \Pi_A, \mathcal V) \quad {\rm and} \quad \liminf_{n\to \infty} \mathcal F(S^n) = \inf\left\{ F(S): S\in \mathcal S(\A, \Pi_A, \mathcal V)\right\}.$$ 

\paragraph{Compactness for curves.} Since for any $\A,\Pi_A,\mathcal V$ satisfying \eqref{eq:apiv} it is possible to construct a spheroid with area $\A$, volume $\mathcal V$, and divide its surface in two domains such that one has area $\Pi_A$, the infimum above is finite, and there exists $\Lambda>0$ such that $\mathcal F(S^n)\leq \Lambda$ for all $n\in \N$. In \cite{ChoksiVen} we studied the compactness properties of a sequence of surfaces for which the bound 
\begin{equation}
\label{eq:sffbound}
	\int_\S (k_1^2 +k_2^2 )\, dS \,  \leq\,  C
\end{equation}
holds uniformly. Owing to Lemma \ref{lemma:fundest}, we can apply the results in \cite[Lemma 2.4, Lemma 2.5, Lemma 3.6, Proposition 3.7]{ChoksiVen} to the sequence of surfaces $\{ \Sigma_m^n\}$. 
 We summarize these results in the following proposition. 
Regarding notation, we denote the first and the second component of a curve $\g_j$ by $(\g_j)_1$ and $(\g_j)_2$\,.
\begin{prop}
\label{prop:compcurv}
Let $S^n$ be a sequence of finite systems of admissible surfaces and phases satisfying the bound \eqref{eq:sffbound} for some fixed constant $C>0$. Denote the generating curves by $(\g_1^n,\ldots,\g^n_{m(n)})$. Then, there exists a subsequence $S^{n_k}$ such that
\begin{itemize}
	\item[(i)] the cardinality $m(n_k)$ of the system is uniformly bounded. Therefore, it is not restrictive to assume that $m(n_k)\equiv \omega$, for some constant $\omega>0$.
	\item[(ii)] There exists a system of curves $\g_1,\ldots,\g_{J}$ such that for all $j=1,\ldots,J$ ($J\leq \omega$)
		$$\g_j \in W^{1,\infty}((0,1);\R^2_+),\qquad (\g_j)_1>0\quad \text{a.e. in } (0,1),\qquad \ddot\g_j \in L^2(\mu_{\g_j};\R^2_+),$$ 
		and, up to a permutation of the indices,
		\begin{align} 
			\g_j^{n_k} &\to \g_j\qquad \quad\text{strongly in }H^1((0,1);\R^2_+),\label{eq:convh1}\\
			\ddot\g_j^{n_k} &\weakto \ddot\g_j\qquad \quad\text{weakly in }L^2(\mu_{\g_j^{n_k}};\R^2_+).
		\end{align}	
	\item[(iii)] For $j=J+1,\ldots,\omega$, the curves $\g_j^{n_k}$ converge to a point lying on the $z$-axis, i.e. 
	$$ (\g_j^{n_k})_1 \to 0,\qquad (\g_j^{n_k})_2 \to z_j\,,\qquad \text{strongly in }H^1((0,1))$$
	for some $z_j\in \R$.
	\item[(iv)] For $j=1,\ldots,J$, it holds $\#\{(\g_j)_1=0\}<+\infty$ and $\g_j\in W^{2,2}(\{(\g_j)_1>0\};\R^2_+)$	
	\item[(v)] For $j=1,\ldots,J$, each curve $\g_j$ meets the $z$-axis orthogonally, so, in particular, the generated surface $\S_j$ is the union of a finite number of $C^1$-regular surfaces.
	\item[(vi)] The area and volume constraints pass to the limit, i.e.,
	\begin{align*}
		\lim_{n\to \infty}\sum_{j=1}^{m(n)}|\S_j^n| =\sum_{j=1}^{J}|\S_j|,\qquad \qquad
		\lim_{n\to \infty}\sum_{j=1}^{m(n)}\V(\S_j^n) =\sum_{j=1}^{J}\V(\S_j).
	\end{align*}	
\end{itemize}	 
\end{prop}
Since singularities for $\dot \g_j$ can occur only on the $z$-axis, i.e. where $(\g_j)_1=0$, points (iv) and (v) imply that the limit system can be reparametrized as a finite family of admissible curves \cite[Corollary 2.9]{ChoksiVen}. 

\paragraph{Compactness for phases.} We turn now to the question of compactness for the phases. Since the system $S^n$ has fixed cardinality, it is enough to study the behavior of a single couple. Let $(\S^n,\vfi^n)$ be a sequence of admissible surface-phase couples, where $\S^n$ is generated by $\g^n$, and let $\g$ be a curve such that $\g^n\to \g$ as in Proposition \ref{prop:compcurv} (ii), (iv). 
First of all, since $\vfi^n$ takes values in $\{0,1\}$, we can find a subsequence (not relabeled) and a function $\vfi\in L^\infty(0,1)$ such that
\begin{equation}
\label{eq:convlinf}
	\vfi^n \weaksto \vfi\qquad \text{weakly-* in }L^\infty(0,1),\qquad 0\leq \vfi(t)\leq 1\quad \text{for a.e. }t\in(0,1).
\end{equation}
 Since convergence \eqref{eq:convh1} implies that $\g_1^n \to \g_1$ uniformly in $C^0([0,1])$, for every compact $K\subset \{\g_1>0\}$ there exists $\e>0$ such that $\g_1^n(t)\geq \e$ for all $t\in K.$ More precisely, since by Proposition \ref{prop:compcurv}-(iv) the set $\{\g_1=0\}$ is finite, for every $\d >0$ there exists a compact set $K=K(\d) \subset \{\g_1>0\}$ and a positive number $\e=\e(\d)>0$ such that
 \begin{equation}
 \label{eq:kd}
 	\mathcal L^1([0,1]\backslash K)\leq \frac\d2\qquad \text{and}\qquad \g_1^n(t)\geq \e\quad \forall\,t\in K.
 \end{equation}
 Owing to Lemma \ref{lemma:fundest}
$$\Lambda \geq \F(\S^n,\vfi^n)\geq C\left(\int_{\S^n}(k_1^n)^2+(k_2^n)^2dS -|\S^n|\right)+\s\hf^1(\Gamma^n).$$
Recalling \eqref{eq:defline} we obtain
$$ \frac{\Lambda +C\A}{2\pi\s}\geq \frac{1}{2\pi}\hf^1(\Gamma^n) = \sum_{t\in J(\vfi^n)}\g^n_1(t) \geq \sum_{t\in J(\vfi^n)\cap K}\g^n_1(t)\geq \e \hf^0(J(\vfi^n)\cap K) .$$
Let $U\subset K$ be an open set such that $\mathcal L^1(K \backslash U)\leq \delta/2$, we have that $\vfi^n\in BV(U)$ and 
\begin{align*} 
	\|\vfi^n\|_{BV(U)}&=\|\vfi^n\|_{L^1(U)}+\|D\vfi^n\|(U)\\
					&=\int_U \vfi^n(t)\, dt + \hf^0(J(\vfi^n)\cap U)\leq 1 +\frac{\Lambda +C\A}{2\pi\s\e}.
\end{align*}
By classical compactness and lower semicontinuity for $BV$ functions (see Theorems \ref{th:thlsc} and \ref{th:thcpt}), there exists a subsequence $\vfi^{n_k}$ and a function $\bar\vfi\in BV(U)$ such that
\begin{align}
	&\vfi^{n_k} \to \bar\vfi\qquad \text{strongly in }L^1(U)\label{eq:phstrong},\\
	&\liminf_{n_k\to \infty}\|D\vfi^{n_k}\|(U)\geq \|D\vfi\|(U).
\end{align}
By \eqref{eq:convlinf}, we also have that $\bar \vfi =\vfi$ a.e.\@ in $U$ and thus $\vfi$ is the strong limit for the whole sequence $\vfi^n$.  By \eqref{eq:kd} and \eqref{eq:phstrong}, we found that for every $\delta>0$ there exists an open set $U\subset (0,1)$ and a function $\bar \vfi$ such that
\begin{align*} 
	\lim_{n \to \infty}\int_0^1 |\vfi^{n}-\vfi|\, dt &= \lim_{n \to \infty}\left(\int_U |\vfi^n-\vfi|\, dt + \int_{[0,1] \backslash U} |\vfi^n-\vfi|\, dt \right)\\
		&\leq \lim_{n \to \infty}\left(\int_U |\vfi^n-\bar \vfi|\, dt + 2\delta \right)\\
		&\leq 2\delta.
\end{align*}	
Since $\delta$ is arbitrary, we obtain strong convergence in $L^1(0,1)$. Recalling that strong convergence, up to subsequences, implies convergence almost everywhere,  we conclude that convergence \eqref{eq:convlinf} is improved to
$$ \vfi^n \to \vfi\qquad \text{strongly in }L^1(0,1),\qquad \vfi\in BV_{\textup{loc}}(\{\g_1>0\};\{0,1\}),$$
or, actually, noting that  $|\vfi^n|\leq 1$, to
$$ \vfi^n \to \vfi\qquad \text{strongly in }L^p(0,1),\qquad \vfi\in BV_{\textup{loc}}(\{\g_1>0\};\{0,1\}),$$
for all $p\in [1,\infty).$ Note that $L^1$-convergence for the phases, combined with convergence \eqref{eq:convh1} for the curves, implies that the phase area constraint passes to the limit: if $S^n=\left\{(\S^n_i,\vfi^n_i))\right\}_{i=1,\ldots,\omega}\in \mathcal S(\A,\Pi_A,\mathcal V)$, then
$$ \Pi_A=\sum_{i=1}^\omega 2\pi\int_0^1\vfi_i^n(t)(\g_i^n)_1(t)|\dot \g_i^n|dt\ \stackrel{n\to \infty}{\to}\ \sum_{i=1}^\omega 2\pi\int_0^1\vfi_i(t)(\g_i)_1(t)|\dot \g_i|dt,$$
and since $(\g_i)_1\equiv 0$ for $i=J+1,\ldots,\omega$, 
$$\sum_{i=1}^J 2\pi\int_0^1\vfi_i(t)(\g_i)_1(t)|\dot \g_i|dt=\Pi_A.$$
\smallskip

We collect these computations and the results of Proposition \ref{prop:compcurv} in the following 
\begin{prop}[Compactness]
\label{prop:comp}
Let the area, $A$-phase area and volume constraints $\A,\Pi_A,\mathcal V$ be given. Let 
$$S^n=\left\{(\S^n_1,\vfi^n_1),\ldots,(\S^n_{m(n)},\vfi^n_{(m(n))})\right\}$$ 
 be a sequence of systems of admissible surfaces-phases such that $S^n\in \mathcal S(\A, \Pi_A, \mathcal V)$ and 
$ \mathcal F(S^n)\leq \Lambda,$ for some constant $\Lambda>0$. Then there exists a subsequence $S^{n_k}$ and an admissible system
$$S=\left\{(\S_1,\vfi_1),\ldots,(\S_J,\vfi_J)\right\} \in \mathcal S(\A, \Pi_A, \mathcal V)$$
such that for all $j=1,\ldots,J$ 
$$\g_j \in W^{1,\infty}((0,1);\R^2_+),\qquad (\g_j)_1>0\quad \text{a.e. in } (0,1),\qquad \ddot\g_j \in L^2(\mu_{\g_j};\R^2_+),$$ 
$$ \vfi_j \in BV_{\textup{loc}}(\{(\g_j)_1>0\};\{0,1\})$$
and, up to a permutation of the indices, 
	\begin{align} 
		\g_j^{n_k} &\to \g_j\qquad \quad\text{strongly in }H^1((0,1);\R^2_+),\label{comp:1}\\
		\ddot\g_j^{n_k} &\weakto \ddot\g_j\qquad \quad\text{weakly in }L^2(\mu_{\g_j^{n_k}};\R^2_+),\label{comp:2}\\
		\vfi_j^{n_k} &\to \vfi\qquad \quad \, \text{strongly in }L^p(0,1), \quad \forall\,p\in [1,\infty).\label{comp:3}
	\end{align}
and
	\begin{equation}
	\label{comp:4}		
		\liminf_{n_k\to \infty}\|D\vfi^{n_k}\|(U)\geq \|D\vfi\|(U)\qquad \forall\,U\subset\subset\{\g_1>0\}.	
	\end{equation}	
\end{prop}


\subsection{Lower semicontinuity}
In this Section we prove the following 
\begin{prop}[Lower semicontinuity]
\label{prop:lsc}
Let $(\S^n,\vfi^n)$ be a sequence of admissible couples of surfaces-phases, converging to the couple $(\S,\vfi)$ in the sense of  \eqref{comp:1}--\eqref{comp:3}, then 
$$
	\liminf_{n \to \infty} \F(\S^n,\vfi^n) \geq \F(\S,\vfi).
$$
\end{prop}
\begin{proof}
We first address the lower semicontinuity of the term $\s\hf^1(\Gamma)$. We need to prove that if the couple $(\g^n,\vfi^n)$ converges to $(\g,\vfi)$ as in \eqref{comp:1}--\eqref{comp:3}, then 
\begin{equation}
\label{eq:lscpre}
	\liminf_{n \to \infty}\s\hf^1(\Gamma^n) = \liminf_{n \to \infty}\s 2\pi \int_0^1 \g_1^n\,d\|D\vfi^n\| \geq \s 2\pi \int_0^1 \g_1\,d\|D\vfi\|= \s\hf^1(\Gamma). 
\end{equation}
This result would be straightforward, \emph{if}  we had $\vfi^n\weaksto \vfi$ in $BV(0,1)$. Since $\vfi^n$ is only $BV_{\textup{loc}}(\{\g_1^n>0\};\{0,1\})$, we need one remark:
\begin{lemma}
\label{lemma:hlsc}
Let $\Omega\subset \R$ be open and bounded. Let $\g,\g^n\in C_0(\Omega)$ be nonnegative, and $\vfi,\vfi^n \in BV_{loc}(\Omega)$ be such that 
\begin{align}
	&\g^n \to \g& &\text{uniformly in }C^0(\Omega), \label{eq:lm2uni}\\
	&\liminf_{n\to \infty}\|D\vfi^n\| (U)\geq \|D\vfi\|(U)\hspace{-1cm} & &\text{for all }U\subset \subset \Omega.\label{eq:lm2bv}
\end{align}	
If there exists $C>0$ such that
\begin{equation}
\label{eq:lmbd}
	\sup_n \int_\Omega \g^n\, d\|D\vfi^n\|\leq C,
\end{equation}
then $\g\|D\vfi\|$ is a finite Radon measure on $\Omega$ and 
$$ \liminf_{n\to \infty}\int_\Omega \g^n\, d\|D\vfi^n\| \geq \int_\Omega \g\, d\|D\vfi\|.  $$
\end{lemma}
\begin{proof}
Let $\psi\in C_c(\Omega)$, with $0\leq \psi \leq 1$,
$$ \int_\Omega \g^n\, d\|D\vfi^n\| \geq \int_\Omega \g^n\psi\, d\|D\vfi^n\| =  \int_\Omega (\g^n-\g)\psi\, d\|D\vfi^n\| + \int_\Omega \g\psi\, d\|D\vfi^n\|. $$	
Since $\vfi^n\in BV_\loc(\Omega)$, there exists $M=M(\psi)$ such that  
$$ \int_\Omega \psi\,d\|D\vfi^n\|\leq M.$$ 
By \eqref{eq:lm2uni}, for all $\e>0$ there exists $\bar n$ such that $\sup_\Omega|\g^n-\g|\leq \e$ for $n\geq \bar n$. We have
\begin{align*}
 	\liminf_{n\to \infty} \int_\Omega \g^n\, d\|D\vfi^n\| &\geq \liminf_{n\to \infty} \left( -\int_\Omega |\g^n-\g|\psi\, d\|D\vfi^n\| + \int_\Omega \g\psi\, d\|D\vfi^n\|\right)\\
		&\stackrel{\eqref{eq:lm2bv}}{\geq} -M\e +\int_\Omega \g\psi\, d\|D\vfi\|.
\end{align*} 
Since $\e$ was arbitrary, we obtain
$$ \liminf_{n\to \infty} \int_\Omega \g^n\, d\|D\vfi^n\| \geq \int_\Omega \g\psi\, d\|D\vfi\|, \qquad\forall\,\psi\in C_c(\Omega),\quad 0\leq \psi \leq 1.$$
Let now $\psi^k\in C_c(\Omega)$ be a sequence such that 
$\psi^k(x)=1$ if $d(\partial \Omega,x)\geq k^{-1}$, $0\leq\psi(x)\leq 1$. Since $\g\in C_0(\Omega)$, there exists a continuous and monotone increasing function $\omega:\Omega \to \R$ such that $\omega(0)=0$ and 
$$ \sup_{x\in \Omega} |\g(x)\psi^k(x)-\g(x)|\leq \sup\left\{ \g(x): d(x,\partial \Omega)\leq \frac1k\right\}\leq \omega\big(  1/k \big),$$ 
i.e., $\g\psi^k \to \g$ uniformly in $C_0(\Omega)$. Since the mapping $\psi\mapsto \int_\Omega \g\psi\,d\|D\vfi\|$ is continuous with respect to uniform convergence, we can extend $\g\|D\vfi\|$ to a finite Radon measure on $\Omega$ and conclude that

$$  \int_\Omega \g\, d\|D\vfi\|=\lim_{k \to \infty} \int_\Omega \g\psi^k\, d\|D\vfi\| \leq \liminf_{n\to \infty} \int_\Omega \g^n\, d\|D\vfi^n\|.$$
\end{proof}

By \eqref{comp:1}--\eqref{comp:4} we can apply Lemma \ref{lemma:hlsc} to a sequence of admissible couples and obtain the lower semicontinuity \eqref{eq:lscpre}.
 \medskip
 
 Regarding the remaining part of the functional, define
\begin{align*}
	\H(\S,\vfi):&= \frac 12\int_\S \kh(\vfi)(H-H_0(\vfi))^2 dS + \int_\S \kg(\vfi)K dS\\
		&=  \pi  \int_0^1 \kh(\vfi)\left( k_1+k_2  -H_0(\vfi)\right)^2 \g_1|\dot \g |\,dt + 2\pi  \int_0^1 \kg(\vfi) k_1 k_2 \g_1|\dot \g |\,dt.
\end{align*}		
We need to show that if the couple $(\g^n,\vfi^n)$ converges to $(\g,\vfi)$ as in \eqref{comp:1}--\eqref{comp:3}, then 
\begin{equation}
\label{eq:hliminf}
	\liminf_{n \to \infty} \H(\S^n,\vfi^n) \geq \H(\S,\vfi). 
\end{equation}
Define the Radon measures
\begin{align*}	
	&\mu_{\g^n}:=2\pi \g^n_1\, |\dot \g^n|\, \L^1\llcorner_{[0,1]}\,,& & \mu_\g:=2\pi \g_1\, |\dot \g|\, \L^1\llcorner_{[0,1]}\,,\\
	&\lambda^n:= \pi \kh(\vfi^n) \g_1^n|\dot \g^n|\L^1 \llcorner_{[0,1]}\,,& & \psi^n:= 2\pi\kg(\vfi^n) \g_1^n|\dot \g^n|\L^1 \llcorner_{[0,1]}\,.
\end{align*}	
By \eqref{comp:3} and by the linearity of $\kh\,,\kg$\,, $H_0$\,,
$$ \kh(\vfi^n) \to \kh(\vfi),\qquad  \kg(\vfi^n) \to \kg(\vfi), \qquad H_0(\vfi^n) \to H_0(\vfi)$$
strongly in $L^p(0,1)$ for every $p \in [1,+\infty)$, while by \eqref{comp:1},
$$ \g_1^n|\dot \g^n| \to \g_1|\dot \g| \qquad \text{uniformly in }C^0([0,1]), $$
so, in particular,  $\lambda^n \weaksto \lambda:=\frac12 \kh(\vfi) \g_1|\dot \g|\L^1 \llcorner_{[0,1]}$ and $\psi^n \weaksto \psi:=\kg(\vfi) \g_1|\dot \g|\L^1 \llcorner_{[0,1]}$ as measures.   Recalling also \eqref{eq:boundkunif}, it is straightforward to check that weak (resp.\@ strong) convergence in $L^p(\mu_{\g^n})$ is equivalent to weak (resp.\@ strong) convergence in $L^p(\lambda^n)$, or in $L^p(\psi^n)$, in the sense of Definition \ref{def:weakags}. By convergence \eqref{comp:2} and \cite[Lemma 3.4]{ChoksiVen},
$$ H^n \weakto H \quad \text{weakly in }L^2(\lambda^n),\qquad K^n \weakto K \quad \text{weakly in }L^1(\psi^n).$$ 
By Theorem \ref{th:ags} and Lemma \ref{lemma:moser} we conclude that
\begin{align*}
	\liminf_{n \to \infty} &\int \frac{\kh}{2}(\vfi^n)(H^n-H_0(\vfi^n))^2d\mu_{\g^n} +\int \kg(\vfi^n)K^n d\mu_{\g^n}  \\
			&= \liminf_{n \to \infty} \int (H^n-H_0(\vfi^n))^2 d\lambda^n +\int K^n d\psi^n \\
			&\geq \int (H-H_0(\vfi))^2d\lambda +\int K d\psi \\
			&= \int \frac{\kh}{2}(\vfi)(H-H_0)^2d\mu_{\g} +\int \kg(\vfi)K d\mu_{\g}.
\end{align*}
This proves \eqref{eq:hliminf}, which together with \eqref{eq:lscpre} concludes the proof of Proposition \ref{prop:lsc}.
\end{proof}

\subsection{Proof of Theorem \ref{th:main}} 
\label{ssec:proof}
Let the total area, $A$-phase area and volume constraints $\A,\Pi_A,\mathcal V$ be given, such that the isoperimetric inequality and the bound on the phase \eqref{eq:apiv} are satisfied. Let the parameters $\kh^i\,,\kg^i$ satisfy \eqref{eq:assk} and $H_0^i\in \R$, for $i=A,B$. Let the set $\mathcal S(\A,\Pi_A,\mathcal V)$ and the functional $\mathcal F$ be given as in the statement of Theorem \ref{th:main}. 

Let $S^n=\big((\S_1^n,\vfi^n),\ldots,(\S_{m(n)}^n,\vfi_{m(n)}^n)\big)\in \mathcal S(\A,\Pi_A,\mathcal V)$ be a sequence of admissible couples of surfaces and phases such that
\begin{equation}
\label{eq:minimizing}
	\lim_{n\to \infty}\mathcal F(S^n)=\inf_{S\in \mathcal S(\A,\Pi_A,\mathcal V)}\mathcal F(S).
\end{equation}		
Since $S^n$ satisfies $\mathcal F(S^n)\leq \Lambda$, for a suitable $\Lambda>0$, by Lemma \ref{lemma:fundest} there is a constant $C>0$ such that 
$$\sum_{i =1}^{m(n)}\left(\int_{\S_i^n} (\k_{1,i}^n)^2+(\k_{2,i}^n)^2\,dA\right) \leq C(\A+\mathcal F(S^n))\leq C(\A+\Lambda).$$ 
We can therefore apply Proposition \ref{prop:comp}, and find admissible couples $(\g_1,\vfi_1),\ldots,(\g_J,\vfi_J)$ and a subsequence (not relabeled) of constant cardinality $m(n)\equiv m$, such that for $0\leq j\leq J$, as $n\to \infty$  
$$ \g_j^n \to \g_j\,,\qquad \vfi^n_j \to \vfi_j$$
in the sense of convergence \eqref{comp:1}, \eqref{comp:3}, while $\g^{n}_{J+1},\ldots,\g^n_m$ shrink to points, thus not contributing to the total area of the system. The system $S$ of surfaces-phases couples generated by $((\g_1,\vfi_1),\ldots,(\g_j,\vfi_j))$ satisfies the total area, phase area and enclosed volume constraints. By the lower semicontinuity Proposition \ref{prop:lsc}
$$ \liminf_{n \to \infty} \mathcal F(S^n)\geq \mathcal F(S),$$
so that, by \eqref{eq:minimizing}, $\mathcal F(S)= \inf \mathcal F$. The proof of Theorem \ref{th:main} is thus complete.

\subsubsection*{Acknowledgements} We would like to thank Eliot Fried for introducing us to this fascinating subject and for many stimulating discussions. M.\,V. acknowledges Luca Mugnai, for suggesting the problem in Figure \ref{fig1}.

\end{document}